# Effect of Sb deficiency on the thermoelectric properties of $Zn_4Sb_3$

Anup V. Sanchela*, C. V. Tomy

*Department of Physics, Indian Institute of Technology Bombay, Mumbai-400076*

Ajay D. Thakur

*Department of Physics, Indian Institute of Technology Patna, Patna-800013, India*
*&*
*Centre for Energy & Enviornment, Indian Institute of Technology Patna, Patna-800013, India.*

**Abstract.** We have investigated the effect of Sb-deficiency on the thermoelectric figure of merit (*zT*) of $Zn_4Sb_3$ prepared by solid state reaction route. At high temperatures, the Seebeck coefficient (*S*) and electrical conductivity ($\sigma$) increase with increase in Sb deficiency whereas the thermal conductivity ($\kappa$) decreases giving rise to an increase in the overall *zT* value. The observations suggest that creation of vacancies could be an effective route in improving the thermoelectric properties of $Zn_4Sb_3$ system. This coupled to nanostructuring strategy could lead to the ultimate maximum value of *zT* in this system for high temperature thermoelectric applications.

**Keywords**: A. Thermoelectric materials; B. Solid state route; B. deficiency; D. Thermoelectric properties.

*Corresponding Author
Email: phonon.sanchela@gmail.com

## 1. Introduction

Over past few decades, thermoelectric materials have attracted a lot of interest due to their potential application in solid state energy conversion devices [1, 2]. The efficiency of thermoelectric devices is determined by the dimensionless figure of merit $zT= S^2T/\rho\kappa$, where $S$ is the Seebeck coefficient, $T$ is the absolute temperature, $\rho$ is the electrical resistivity and $\kappa$ is the thermal conductivity which is the combination of lattice thermal conductivity ($\kappa_{lat}$) and electronic thermal conductivity ($\kappa_e$) [3]. In order to achieve a high thermoelectric performance, a high value of $S$, a low value of $\kappa$ and a low value of $\rho$ are required simultaneously. The compound $Zn_4Sb_3$ belongs to a class of materials classified as "Phonon-Glass Electron-Crystal (PGEC)" which are well known for a good electrical conductivity (electron-crystal) and a very low lattice thermal conductivity (phonon-glass) [4, 5]. A reasonably high $zT$ of 1.3 at 670 K was reported in this material [6, 7]. The parent material $Zn_4Sb_3$ exists in three structurally distinct phases α, β and γ. The transformation from α to β phase occurs at 263 K followed by a transformation from β to γ phase at 765 K. The β phase exists between 263 K and 765 K and exhibits the highest $zT$ [7, 11]. The β phase reveals a hexagonal rhombohedric crystal structure with space group $R\bar{3}c$ with 66 atoms per unit cell [6, 7, 12]. The very low lattice thermal conductivity originates from the disordered three Zn interstitials [4]. In addition soft localized dumbbell vibrations of Sb also play a crucial role in lowering lattice thermal conductivity [8]. Phonon scattering, resulting from point defects caused by introducing Sb has the potential to further decrease the lattice thermal conductivity. Various substitutions have been tried at Zn and Sb sites to further improve the thermoelectric properties. However we have not come across any reports on Sb site deficiency in $Zn_4Sb_3$ compound.

The prior works mainly focused on Zn interstitial sites and their disordering [4, 5]. Here, we create vacancy by introducing deficiency at Sb site in $Zn_4Sb_3$ compound leading to missing interatomic linkages [9] thereby lowering $\kappa_{lat}$ [10].

## 2. Experimental details

Polycrystalline samples of $Zn_4Sb_{3-x}$ (x = 0.0, 0.1, 0.2) were prepared by the solid state reaction route. Shots of Zn (99.99%) and powder of Sb (99.99%) were mixed in stoichiometric proportion, ground together and heated in an evacuated ($10^{-5}$ Torr) quartz tube up to 900°C at a heating rate of 45°C/h. After soaking the charge at this temperature for 2h, the quartz tube was quenched in cold water similar to recipes reported for $Zn_4Sb_3$ [7]. The ingots (light grey in colour) thus obtained were used for further measurements. Powder X-ray diffraction (XRD) patterns were obtained using a Philips X'Pert PRO X-ray diffraction system using Cu-$K_\alpha$ radiation. The Raman scattering measurements were carried using a Jobin-Yvon T6400 Raman system in the micro-Raman configuration. An argon (Ar+) ion laser with wavelength 514.5 nm was used as the excitation source. UV measurements were carried out by using a LAMBDA 1050 UV/Vis/NIR spectrometer along with 150-mm integrating sphere (PerkinElmer, Inc., Shelton, CT USA). Measurements of the thermal properties were carried using the thermal transport option (TTO) of Physical Property Measurement System (PPMS), Quantum Design (USA) utilizing the two probe configuration in the temperature range from 5 K to 395 K. Four probe resistivity measurements were performed using the resistivity option of the PPMS in the temperature range from 5 K to 395 K. High temperature Seebeck and resistivity measurements were carried using the thermoelectric system ZEM-3, ULVAC-RIKO (Japan) in the temperature range from 345 K to 632 K.

## 3. Results and discussion

Figure 1 shows the X-ray powder diffraction patterns at 300 K for the ingots of $Zn_4Sb_{3-x}$ (x = 0.0, 0.1, 0.2) samples. All the diffraction peaks could be indexed in the Rietveld analysis by assuming a hexagonal rhombohedric crystal structure with space group $R\bar{3}c$ [7, 12]. The lattice parameters, relative density and the theoretical (X-ray) density obtained from the fit are listed in table 1. The XRD patterns illustrate that the dominant phase in all the samples is of $Zn_4Sb_3$ (a small amount of ZnSb impurity phase was found in some of the samples which are marked with an asterisk in the figure) [13]. The lattice parameters obtained from our fit are in good agreement with those reported in the literature [14]. The densities of the ingots were calculated from the mass and dimensions of the parallelepiped shaped samples.

Figure 2 shows plot of $(\alpha h\nu)^2$ versus photon energy hν. The measured optical data was converted to Kubelka-Munk function $\alpha = (1-R)^2/2R$. By suitable extrapolation of the curve, the optical band gap was estimated. The estimated band gap values for $Zn_4Sb_{3-x}$ (x = 0.0, 0.1, 0.2) are 1.29 eV, 1.23 eV and 1.25 eV respectively which is in good agreement with reported band gaps [12, 15, 7, 21].

Figure 3 shows the Raman spectra for $Zn_4Sb_{3-x}$ (x = 0.0, 0.1, 0.2) in the region between 100 - 220 $cm^{-1}$. The major peaks marked (1 & 2) 148 $cm^{-1}$ and 168 $cm^{-1}$ corresponds to the $Zn_4Sb_3$ phase. These are associated with Sb-Sb bondings and Zn-Sb bondings respectively [16, 17]. We have also noticed impurity ZbSb shoulder about at 110 $cm^{-1}$ [18, 21, 26]. Reported $Zn_4Sb_3$ modes are observed at 155 $cm^{-1}$ [16, 19], 152 $cm^{-1}$ [20], 151 $cm^{-1}$ [17], 144 $cm^{-1}$ [21] and 172

$cm^{-1}$ [16, 19, 17] 175 $cm^{-1}$ [20]. The two Raman modes were fitted using Lorentzian profiles. The peaks are observed to be shifted towards lower energies compared to reported modes. It might be due to slightly different micro structures arising at different quenching temperatures. Here we quenched our samples at 900 ℃ instead of 750 ℃ as reported by caillat at al. This was done to obtain a better homogeneity [7]. For $Zn_4Sb_{2.9}$ first Raman mode (peak 1) position is almost same as compared to pure $Zn_4Sb_3$ first Raman mode (peak 1). Second Raman mode of $Zn_4Sb_{2.9}$ and $Zn_4Sb_{2.8}$ both the Raman modes (peak 1and peak 2) are shifted to higher energies with associated broadening while increasing Sb deficiency. The corresponding shift is related with the lattice parameter and unit cell volume, which indicates that Sb deficiency leads to the unit cell contraction (see table 2).

We first show the effect of Sb-deficiency on the electrical resistivity (cooling curves) of $Zn_4Sb_{3-x}$ (x = 0.0, 0.1, 0.2) compounds in Figure 4. The resistivity of stoichiometric $Zn_4Sb_3$ compound shows the expected behavior with a sharp discontinuity at the $\alpha \leftrightarrow \beta$ structural transition [14, 25, 27]. The effect of this first order transition is more evident in the upper inset, where we show the thermal hysteresis between the cooling and heating data. With the Sb-deficiency, we observe a strange behavior in the resistivity; in the $\alpha$-phase resistivity increases while it decreases in the $\beta$-phase from 4.0 mΩcm for stoichiometric compound to 2.7 mΩcm for Sb deficient compounds at 394 K. With a further increase in temperature, the resistivity increases and reaches a maximum of 4.3 mΩcm for $Zn_4Sb_{2.9}$ at 632 K as seen in Figure 4. Note that the low temperature and high temperature measurements were performed using different equipment and they are suitable labeled in Figure 4. Thermal hysteresis between the cooling and heating data across the structural transition was also observed in the Sb-deficient compounds (upper inset in Fig. 4). The anomaly at the structural transition was further investigated through the heat capacity measurement which is shown in Figure 5. The inset panel in Fig. 5 shows the temperature variation of heat capacity in the entire temperature range for the stoichiometric compound and the heat capacities near the structural transition for the Sb-deficient compounds. The heat capacity shows sharp peak across the structural transition, as evident in the main panel where we show the heat capacities on an expanded scale in the close vicinity of the transition. We also observed a small shift in the transition temperature along with an over-all increase in the entropy change for the Sb-deficient compounds, consistent with sharper transitions in the resistivity behavior across the structural transition.

In order for the material to be useful for thermoelectric applications, a decrease in thermal conductivity is also desirable along with a decrease in electrical resistivity. In Figure 6, we show the temperature dependence of κ for $Zn_4Sb_{3-x}$ (x = 0.0, 0.1, 0.2) samples; with the total thermal conductivity (κ) shown in the main panel and the lattice contribution $\kappa_{lat}$ in the right inset and the electronic contribution $\kappa_e$ in the left inset. In order to obtain the lattice thermal conductivity, the electronic part of thermal conductivity was calculated from the resistivity data using the Wiedemann-Fanz law, and subtracted from the total thermal conductivity. The lattice thermal conductivity of all the samples shows an expected behavior (a power law at low temperatures, and a 1/T decrease at high temperatures [22, 23]) until the $\alpha \rightarrow \beta$ phase transition occurs, where the thermal conductivity shows a sharp drop. We see a clear decrease in the lattice thermal conductivity at high temperatures (i.e., above the structural transition) for Sb-deficient samples, even though the electrical conductivity shows an increase that is always beneficial for practical applications. As high temperature thermal conductivity measurements could not be performed using the ULVAC RIKO system, we have extrapolated $Zn_4Sb_{2.9}$ thermal conductivity data for $T$ > 400 K and used those values to estimate $zT$ at $T$ > 400 K. It is quite possible that the defects and vacancies created by the Sb-deficiency cause stronger phonon scattering that would lead to a lower thermal conductivity in our samples and thus the estimated values of *zT* act as a good lower bound.

Results of thermopower measurements are shown in Figure 7 for $Zn_4Sb_{3-x}$ (x = 0.0, 0.1, 0.2) samples. The positive values of the Seebeck coefficient for all the samples mean that the major charge carriers in all the samples are holes. The sudden change in Seebeck coefficient around 250 K is consistent with the structural transition. At high temperatures, the Seebeck coefficient of $Zn_4Sb_{2.9}$ increases and reaches a maximum of 174 μV/K at about 632 K. A slight difference in the values of Seebeck coefficient is observed at T ~ 390 K for measurements performed using the two different systems. We do not see any significant change in the Seebeck coefficient up to 400 K in the $Zn_4Sb_{2.9}$ sample as compared that of $Zn_4Sb_{3.0}$ even though its value is considerable lower for the $Zn_4Sb_{2.8}$ sample.

Collating all the data, we can now calculate *zT* values for our samples, which are shown in Figure 8, as a function of temperature. We have achieved a *zT* value ~ 0.32 at 395 K for the Sb-deficient sample, $Zn_4Sb_{2.9}$ sample which 1.6 times larger that of the stoichiometric sample $Zn_4Sb_{3.0}$ , 1.5 times larger that of $Zn_4Sb_{3.0}$ at 300 K that is reported by Qin et al [24] (see table 3). This enhancement in *zT* is even higher across the structural transition, as can be clearly seen in the figure 8. At the higher temperatures maximum zT reaches about ~ 0.56 in the Sb deficient $Zn_4Sb_{2.9}$ sample at approximately 550 K.

## 4. Conclusions

We have successfully synthesized Sb deficient polycrystalline samples $Zn_4Sb_{3-x}$ (x = 0.0, 0.1, 0.2) by solid state reaction technique and thermoelectric properties have been investigated from 5 K to 632 K. It seems that 10 % ($Zn_4Sb_{2.9}$) Sb deficiency is most effective to enhance *zT* in this system. We could achieve a *zT* value of ~ 0.32 at 395 K which is about 37% higher than ingot $Zn_4Sb_3$ and at high temperature achieve a *zT* value of ~ 0.56 at about 632 K. A 10% deficiency does not seem to reduce the electrical conductivity at 400 K, but does increase the Seebeck coefficient and reduce the thermal conductivity, resulting in improved zT values. Further improvement in *zT* is possible in (i) Sb deficient single crystalline materials and (ii) ball milled and hot pressed samples of Sb deficient polycrystals. These are proposed for possible further studies.

**Acknowledgments**

Authors would like to acknowledge the Indian Department of Science and Technology for partial support through the project IR/S2/PU-10/2006. ADT would like to acknowledge partial support from the Center for Energy and Environment, IIT Patna.

**References**

1. C. Wood, Rep. Prog. Phys. **51**, 459 (1988).
2. T. M. Tritt, and M.A. Subramanian, MRS BULLETIN**. 31,** 189 (2006).
3. B. C. Sales, MRS BULLETIN. 15 (1998).
4. G. J. Snyder, M. Christensen, E. Nishibori, T. Caillat, and B. B. Iversen, Nature Mater. **3**, 458 (2004).
5. E. S. Toberer, K. A. Sasaki, C. R. I. Chisholm, S. M. Haile, W. A. Goddard III, and G. J. Snyder, Phys. Stat. Sol. (RRL) **1**, 253 (2007).
6. T. Caillat, J. P. Fleurial, and A. Borshchevsky, 15th international Conference on Thermoelectrics, IEEE, 151 (1996).
7. T. Caillat, J. P. Fleurial, and A. Borshchevsky, J. Phys. Chem. Solids **58**, 1119 (1997).

8. W. Schweika, R. P. Hermann, M. Prager, J. Perßon, and V. Keppens, PRL **99,** 125501 (2007).

9. C. Wan, Y. Wang, N. Wang, W. Norimatsu, M. Kusunoki, and K. Koumoto, Sci. Technol. Adv. Mater. **11,** 044306 (2010).

10. A. V. Sanchela, V. Kushwaha, A. D. Thakur, and C. V. Tomy, Advanced Materials Research. **665**, 179 (2013).

11. C. Okamura, T. Ueda, and K. Hasezaki, Materials Transactions **51**, 152 (2010).

12. M. Tapiero, S. Tarabichi, J.G. Gies, C. Noguet, J.P.Zielinger, M. Joucla, J. Loison, M. Robino, and J. Henrion, Solar Energy Mater.**12**, 252 (1985).

13. B. L. Pedersen, H. Birkedal, B. B. Iversen, M. Nygren, and P. T. Frederiksen, Appl. Phys. Lett. **89**, 242108 (2006)

14. G. Nakamoto, T. Souma, M. Yamaba, and M. Kurisuet, Journal of Alloys and Compounds **377**, 61 (2004).

15. V. D. Malčić, Ž. B. Mikočević, and K. Itrić, Technical Gazette 18 **1**, 117 (2011).

16. R. Viennois , M. C. Record , V. Izard , and J. C. Tedenac , Journal of Alloys and Compounds **440**, 22 (2007).

17. R. Carlini, M. M. Carnasciali, F. Soggia , S. Campodonico, and G. Zanicchi, Journal of Alloys and Compounds **588** , 361 (2014).

18. P. Jund, R. Viennois, X. Tao, K. Niedziolka, and J. C. Tédenac , Phys. Rev. B **85**, 224105 (2012).

19. F. Rouessac, and R. M. Ayral, Journal of Alloys and Compounds **530**, 56 (2012).

20. A. Denoix, A. Solaiappan, R. M. Ayral, F. Rouessac, and J. C. Tédenac, Journal of Solid State Chemistry **183**, 1090 (2010).

21. S. Sitthichai, T. Thongtem, S. Thongtem, and T. Suriwong, Superlattices and Microstructures **64,** 433 (2013).

22. D. M. Rowe, CRC handbook of thermoelectric, Boca Raton, FL, CRC Press, 1995.

23. Y. Wu, J. Nyle´n, C. Naseyowma, N. Newman, F. J. Garcia-Garcia, and U. Haussermann, Chem. Mater. **21**, 151 (2009).

24. X.Y. Qin, M. Liu, L. Pan, H. X. Xin, and J.H. Sun, J. Appl. Phys. **109**, 033714 (2011).

25. T. Souma, G. Nakamoto, and M. Kurisu, J. Alloys Compd. **340**, 275 (2002).

26. D. V. Smirnov, D. V. Mashovets, S. Pasquier, J. Leotin, P. Puech, G. Landa, and Y. V. Roznovan, Semicond. Sci. Technol. **9**, 333 (1994).

27. S. Bhattacharya, R. P. Hermann, V. Keppens, T. M. Tritt, and G. J. Snyder, Phys. Rev. B **74**, 134108 (2006).

**Figure Captions**

**Fig. 1.** (Color online) XRD patterns of $Zn_4Sb_{3-x}$ (x = 0.0., 0.1, 0.2).

**Fig. 2.** (Color online) The plot of $(\alpha h\nu)^2$ verses hν of $Zn_4Sb_{3-x}$ (x = 0.0.). The plot of $E_g$ versus x, $Zn_4Sb_{3-x}$ (x = 0.0, 0.1, 0.2.) (inset).

**Fig. 3.** (Color online) The Raman spectra of polycrystalline $Zn_4Sb_{3-x}$ (x = 0.0, 0.1, 0.2) samples at room temperature. Peaks are suitably marked and Lorentzian profiles have been fitted.

**Fig. 4.** (Color online) Temperature dependence of the electrical resistivity for $Zn_4Sb_{3-x}$ (x = 0.0., 0.1, 0.2). Thermal hysteresis for $Zn_4Sb_3$ and $Zn_4Sb_{2.9}$ lower inset and upper inset respectively.

**Fig. 5.** (Color online) Temperature dependence of the heat capacity for $Zn_4Sb_{3-x}$ (x = 0.0., 0.1, 0.2).

**Fig.6** (Color online) Temperature dependence of the total thermal conductivity (main panel), lattice thermal conductivity (right inset) and electronic thermal conductivity (left inset) for $Zn_4Sb_{3-x}$ (x = 0.0., 0.1, 0.2).

**Fig.7** (Color online) Temperature dependence of the Seebeck coefficient for $Zn_4Sb_{3-x}$ (x = 0.0., 0.1, 0.2).

**Fig.8** (Color online) Temperature dependence of the figure of merit, ZTs for $Zn_4Sb_{3-x}$ (x = 0.0., 0.1, 0.2).

**Table Captions**

**TABLE I.** Lattice parameters, relative density and theoretical density for ingot $Zn_4Sb_{3-x}$ (x = 0.0, 0.1, 0.2) samples at room temperature.

**TABLE II.** Cell volume ($\dot{A}^3$), peak positions ($cm^{-1}$) and FWHM ($cm^{-1}$) of Raman modes for ingot $Zn_4Sb_{3-x}$ (x = 0.0, 0.1, 0.2) samples at room temperature.

**TABLE III.** Room temperature thermoelectric properties for ingot $Zn_4Sb_{3-x}$ (x = 0.0, 0.1, 0.2) samples which is produced without hot pressed and nanostructuring approach which is compare with ref. 24 hot pressed pure $Zn_4Sb_3$ compound.

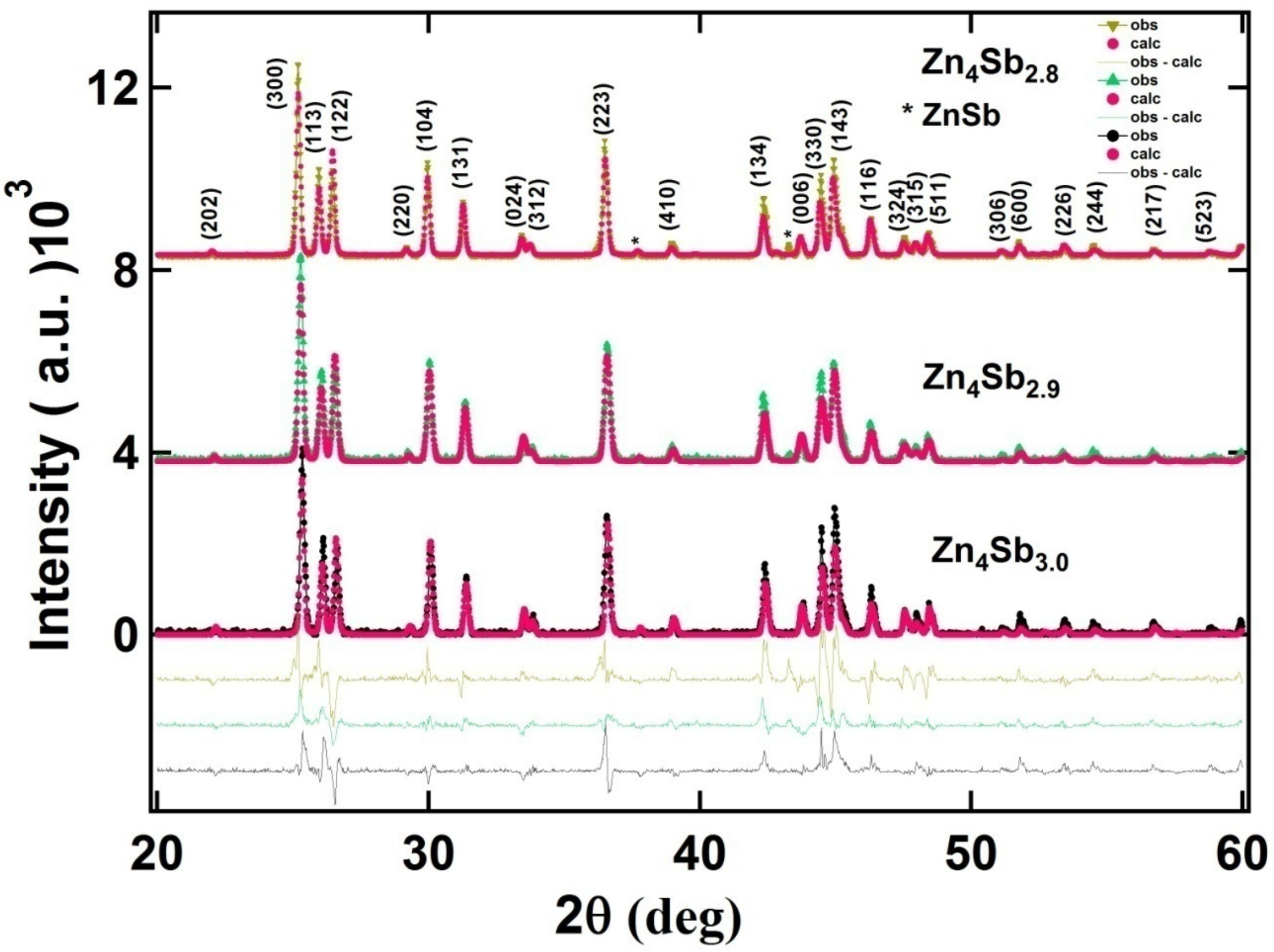

**Figure 1**

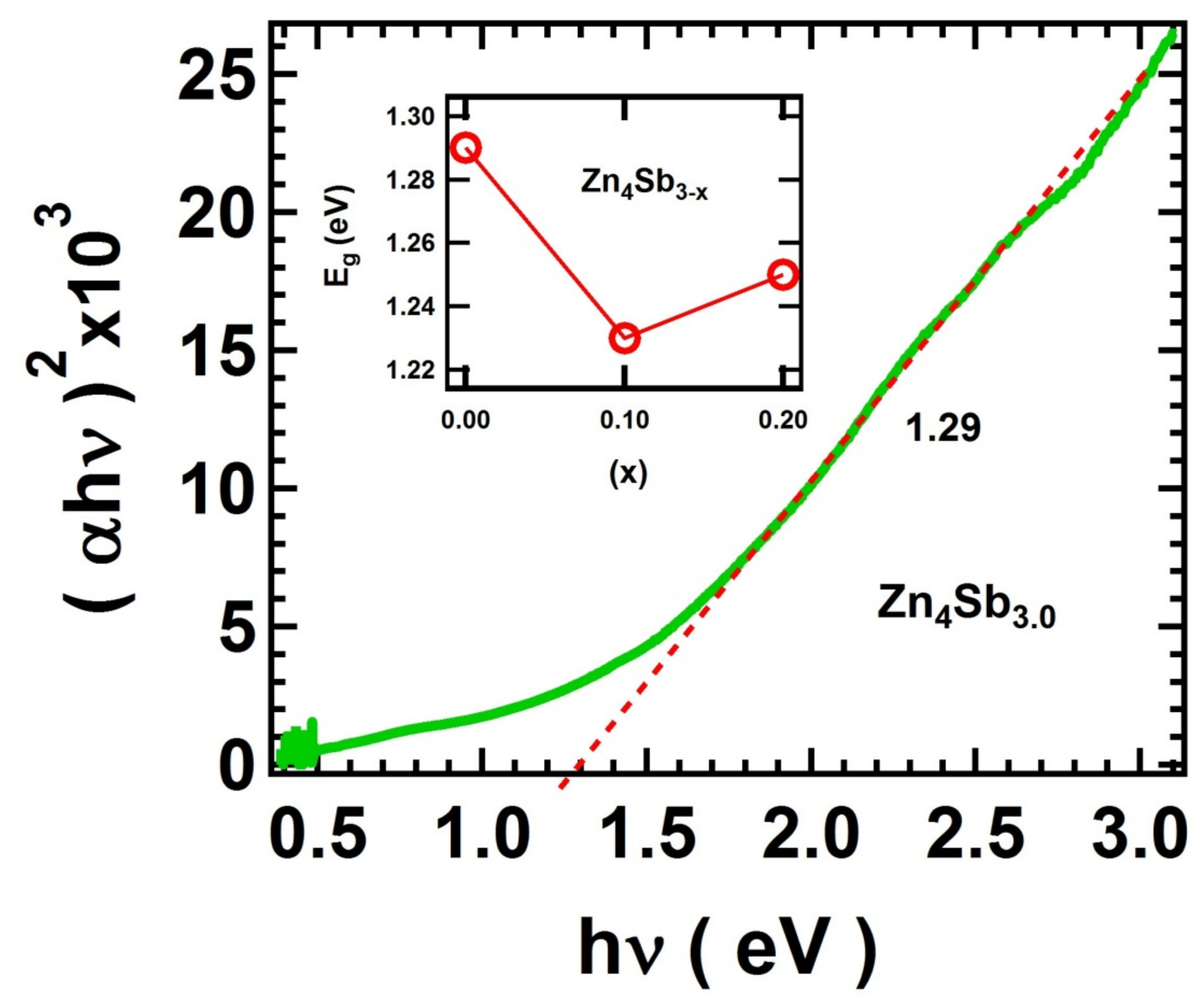

**Figure 2**

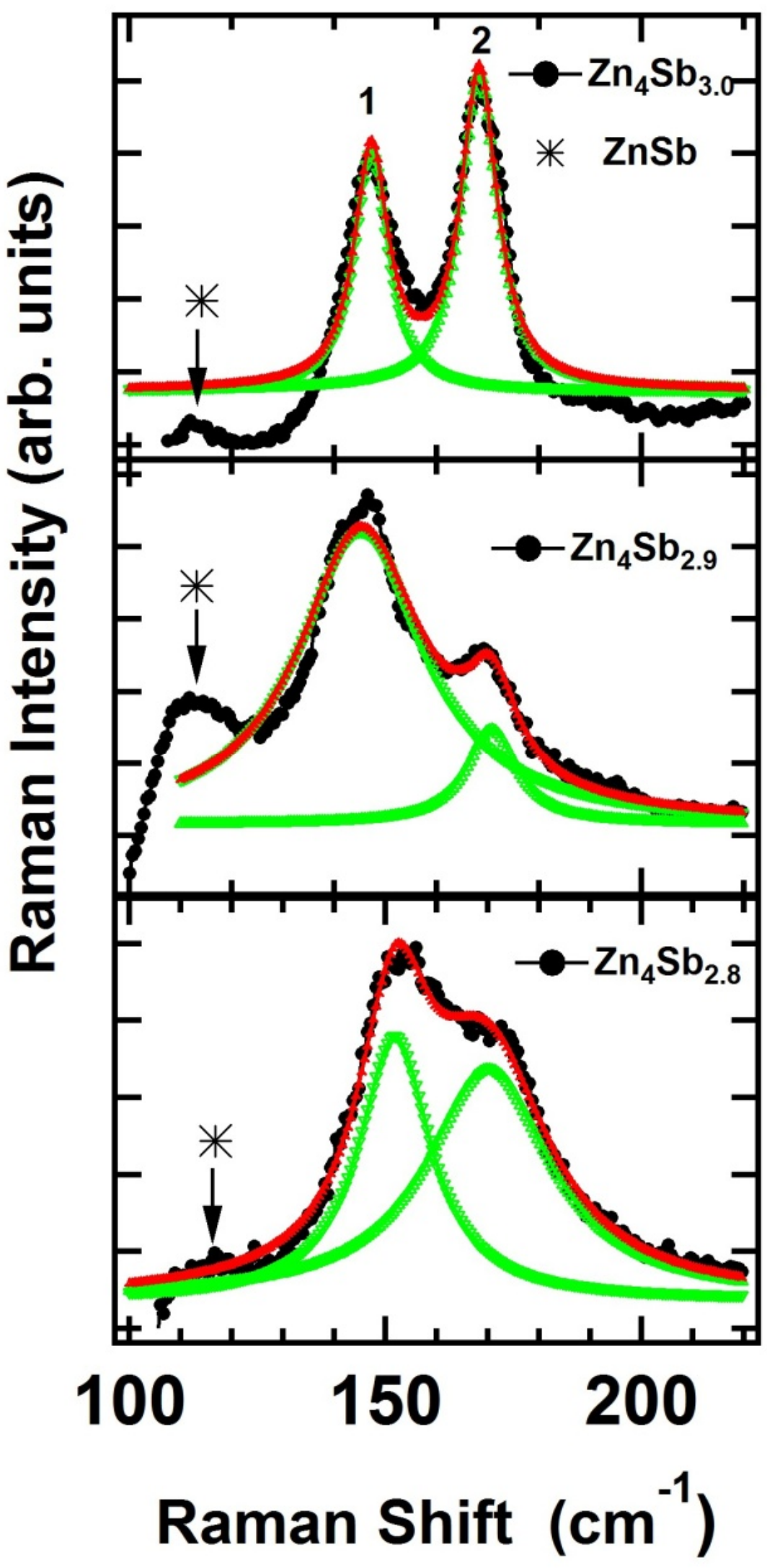

**Figure 3**

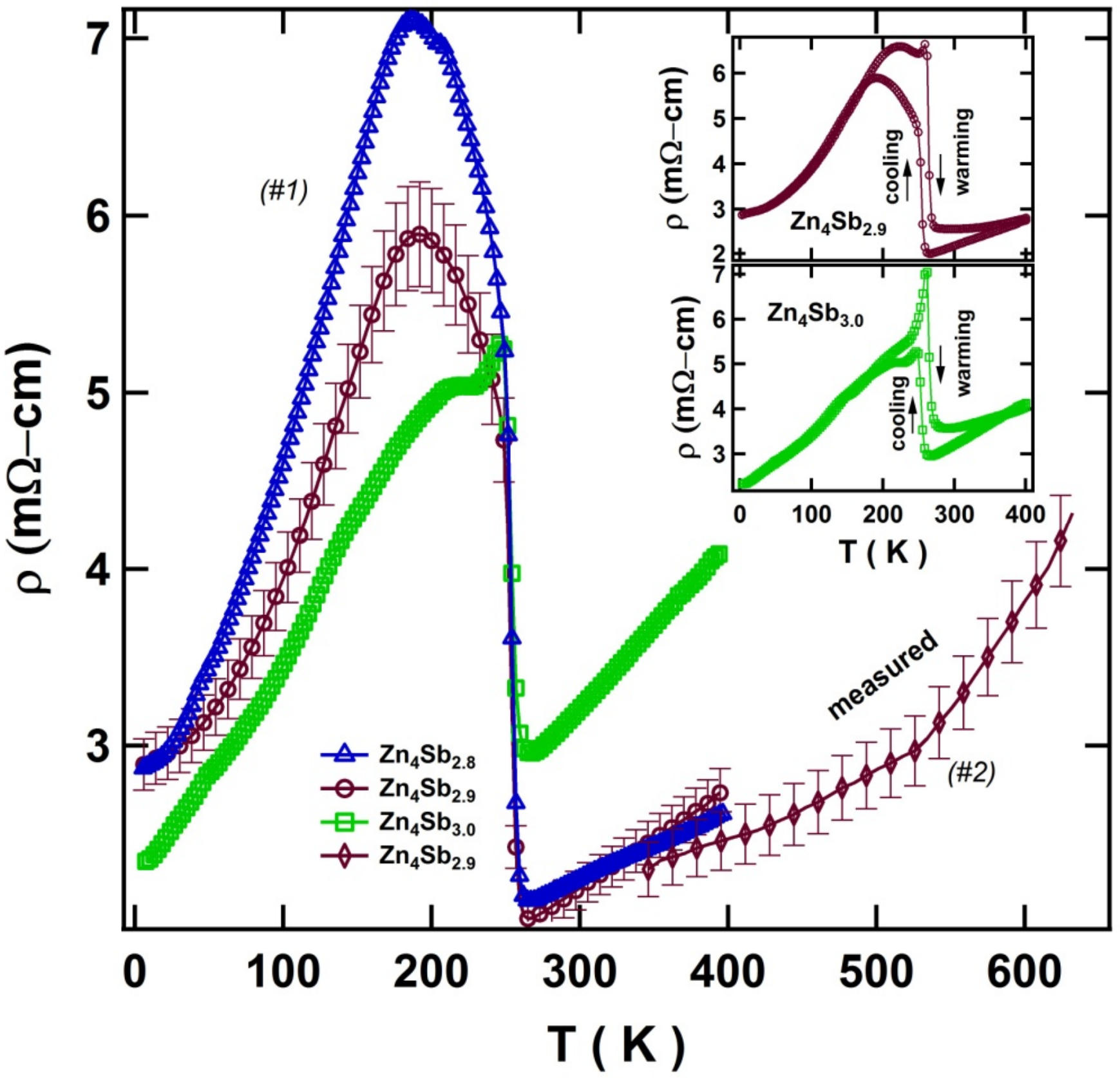

**Figure 4**

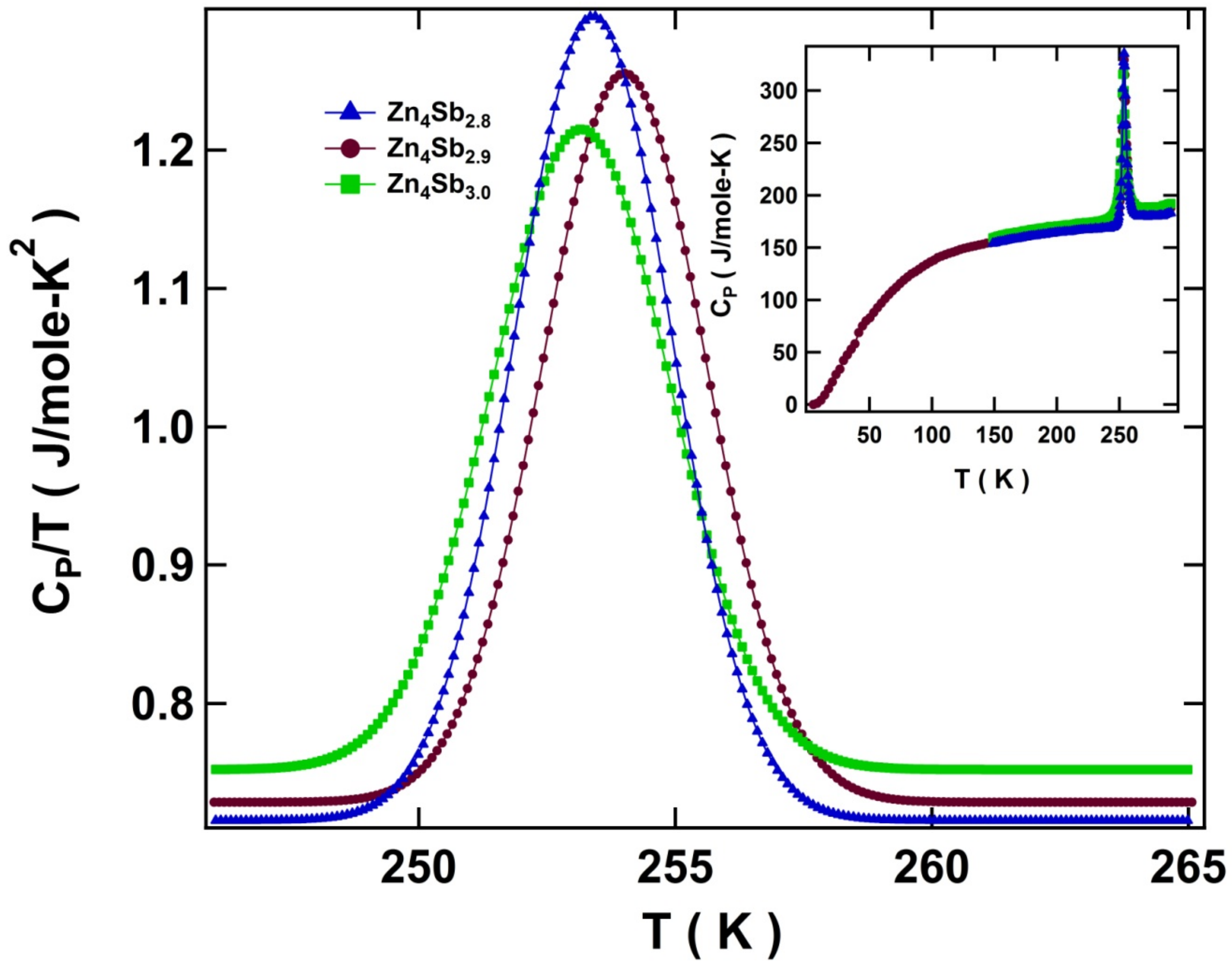

**Figure 5**

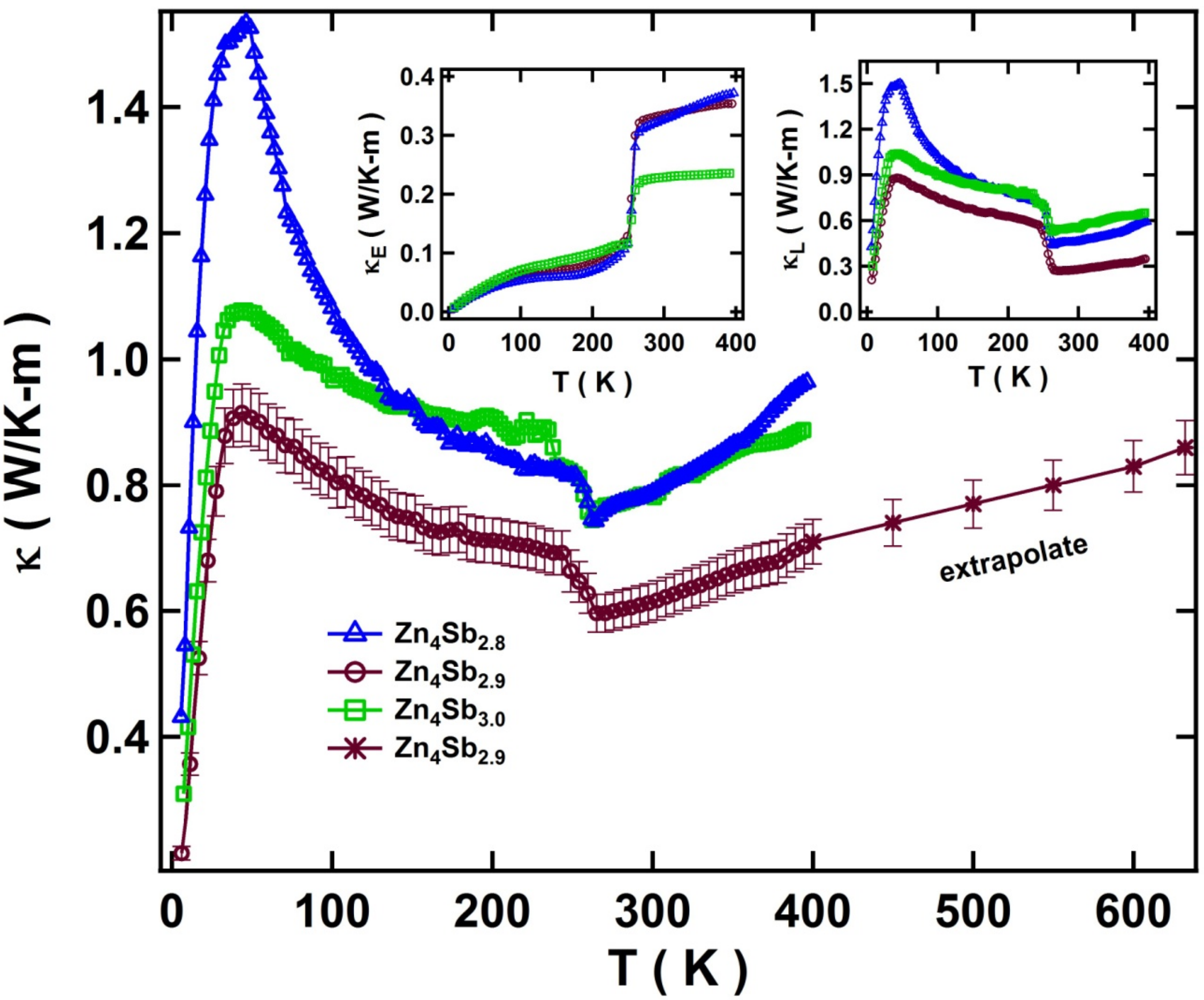

**Figure 6**

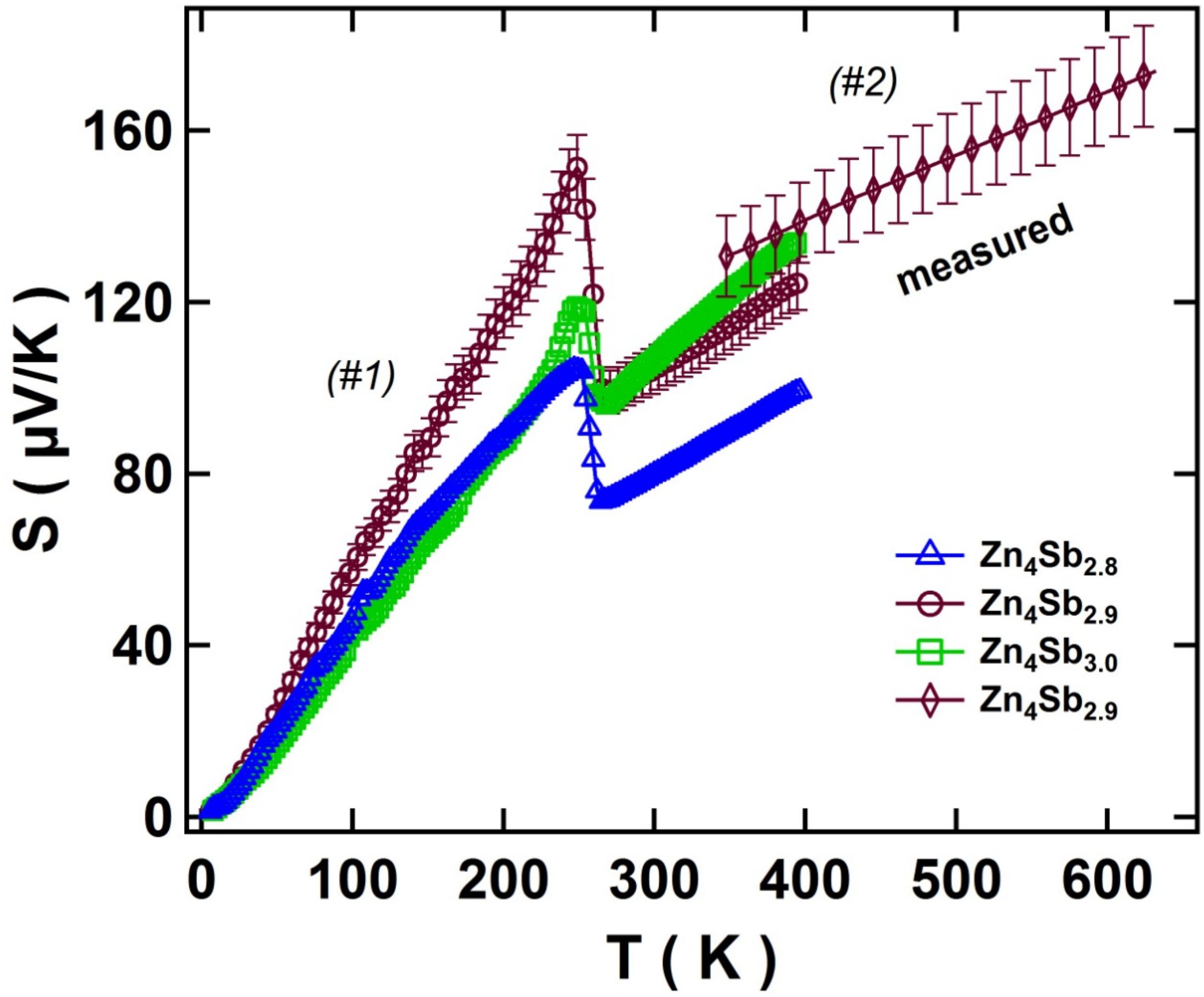

**Figure 7**

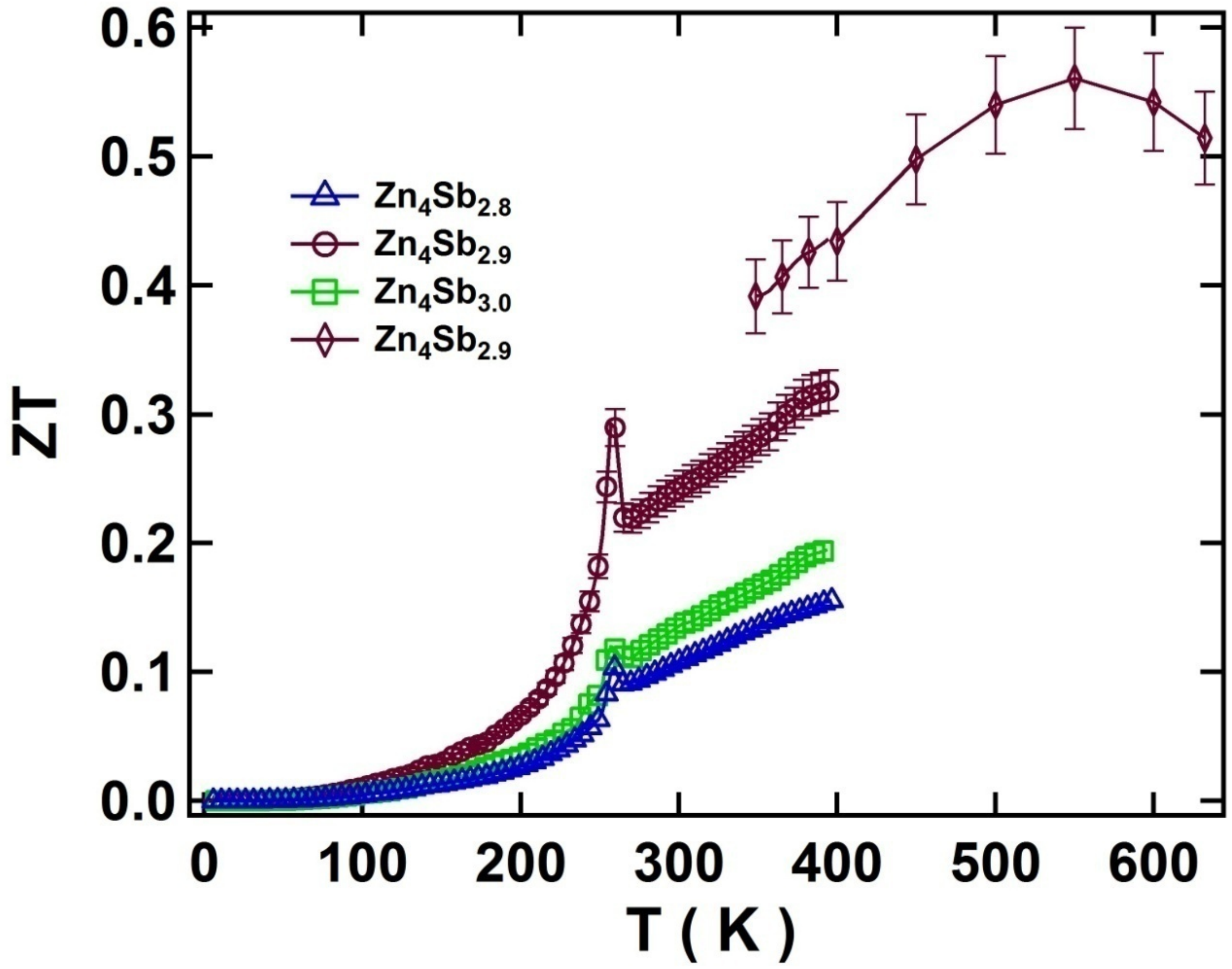

**Figure 8**

# Table I

| | x | a ($\dot{A}$) | c ($\dot{A}$) | Relative Density (%) | Theoretical Density (g/cm$^3$) |
|---|---|---|---|---|---|
| This work Ingot | 0.0 | 12.242 | 12.442 | 82 | 6.122 |
| | 0.1 | 12.237 | 12.435 | 85 | 6.011 |
| | 0.2 | 12.217 | 12.408 | 91 | 5.925 |
| Ref. 14. | 0.0 | 12.239 | 12.440 | 95 | 6.126 |

# Table II

| Compositions | Cell volume | Peak 1 | FWHM | Peak 2 | FWHM |
|---|---|---|---|---|---|
| $Zn_4Sb_{3.0}$ | 1614.95 | 147.35 | 10.22 | 167.79 | 10.00 |
| $Zn_4Sb_{2.9}$ | 1612.80 | 146.49 | 30.44 | 169.19 | 13.28 |
| $Zn_4Sb_{2.8}$ | 1603.90 | 153.69 | 19.38 | 172.51 | 33.21 |

# Table III

| | Compounds | κ (W/Km) | ρ (Ωm) | S (μv/K) | $E_g$ (eV) | ZT |
|---|---|---|---|---|---|---|
| This work Ingot | $Zn_4Sb_{3.0}$ | 0.78 | $3.21\times10^{-5}$ | 106.6 | 1.29 | 0.14 |
| | $Zn_4Sb_{2.9}$ | 0.61 | $2.19\times10^{-5}$ | 104.3 | 1.23 | 0.24 |
| | $Zn_4Sb_{2.8}$ | 0.79 | $3.24\times10^{-5}$ | 79.94 | 1.25 | 0.10 |
| Ref. 24 Hot pressed | $Zn_4Sb_{3.0}$ | 1.25 | $3.32\times10^{-5}$ | 110.0 | ---- | 0.08 |